# Communities of attention networks:
# introducing qualitative and conversational perspectives for altmetrics

Ronaldo Ferreira de Araujo
0000-0003-0778-9561
Lab-iMetrics, Federal University of Alagoas, Brazil
ronaldo.araujo@ichca.ufal.br

**Abstract:** We propose to analyze the level of recommendation and spreading in the sharing of scientific papers on Twitter to understand the interactions of communities around papers and to develop the "Community of Attention Network" (CAN). In this paper, a pilot case study was conducted for the paper 'Pharmacological Treatment of Obesity' authored by Mancini & Halpern (2002), an extensive review of the criteria for evaluating the efficacy of anti-obesity treatments and derived pharmacological agents. The altmetric data was collected from Altmetric.com and the description information for each tweeter was extracted from their Twitter profiles. The data were analyzed with Microanalysis Of Online Data perspective to investigate the formation of a CAN around this focal paper and the context of its formation. The studied article received 736 tweets from 134 different users with a combined exposure of more than 459,018 followers and a high level of spreading (67.26%) and recommendation (28.53%). The user's bios information analysis of who shares the article indicate individual profiles focused on personal issues and strong civic and political engagement. Personal-professional and institutional tweeters of the national political scene are often mentioned in the tweets. In analyzing the content of the tweets, we note that the altmetric score of the paper is a result of its strategic use as an online activism resource and a digital advocacy tool used to mobilize stakeholders for awareness and support activities. This study and the contextual and network perspective it introduces may help to understand the social impact of publications by using altmetrics.

**Keywords:** Altmetrics, Online attention, Communities of attention network, Twitter.

## Introduction

In the context of open access and open science it is envisaged the widespread use of digital resources for supporting and enhancing research activities. It is also expected that practices and attitudes that are more democratic, transparent and collaborative, will amplify the visibility, access and social impact of scientific outputs. There is a movement to encourage the scientific community to publish preferentially in open access journals (Olijhoek, 2011) and to disseminate their partial and final research results in open platforms, institutional repositories, blogs, social media, academic social networks, etc. The promotion of public access to published research in the most widespread manner possible is the central mission of any credible open access journal editor or group of editors.

Online attention indicators have been debated in the context of altmetric studies, which focus on understanding the social impact of research results on the social web. In parallel to the studies that analyze the quantitative aspects of the data – the "how manys" – altmetric approaches should use rich semantic data to investigate the "how" or the "why" (Priem et al., 2010).

From this perspective, some studies set out to analyze the context in which scientific papers are shared online, contributing to the understanding of the reasons for the online attention these papers receive (Nelhans and Lorentzen, 2015; Araújo and Furnival, 2016). Going beyond the "how" and the "why" type of studies, others have analyzed the "who" as an important element for a qualitative understanding of the data in the identification of users and groups of users that share scientific papers (Haustein and Costas, 2015).

Studies developed with these more contextual approaches are growing in the literature. They signal the concern in the altmetric field to contribute to the deeper analysis and investigation of *where* and *how* articles are used by diverse communities that interact with the articles online.

The main research question that this paper tackles is: *How are communities of online attention networks configured around scientific papers?* More specifically, the following two sub-questions are also proposed: *Who shares scientific papers and to whom are scientific papers recommended on Twitter? What is the interaction context of these shares and recommendations?*

Because of its informational and conversational nature, Twitter is a central tool in the research of this kind of research questions. Studies aiming to answer such questions should take a qualitative and social perspective of altmetrics and may contribute to unraveling aspects of the interest of the academic and general public in the subject being discussed in the shared papers.

These studies have investigated the community of attention that gives precedence to the knowledge of the type of attention and involvement that scientific papers might be receiving in other areas, sectors, political references, post-publication comments from peers, citizens, social groups and civil organizations.

**Communities of attention network on twitter**

One of the central issues associated with altmetrics is the identification of communities engaging with scholarly content on social media (Haustein, Bowman, & Costas, 2015; Sugimoto et al, 2017). It is thus of central importance to understand the uses and users of social media in the context of scholarly communication (Sugimoto et al, 2017).

Social media have a central place in research on behavior and habits on the internet as the digital environments in which Brazilian users spend most of their time when they are online (CGI, 2017). The importance and audience acquired by them can be due to the combination of several aspects: (a) identity and self-presentation, as an environment of profile creation and exposure; (b) its diversity of contents, with information of any nature and on any type of object; and (c) its relational social aspect that allows for varied virtual interactions and links.

All of these aspects presented above are recorded and have cumulative features that can be traced in social media such as Twitter and require studies that are capable of measuring the impacts and the implications of their use in the context of scholarly communication. According to some review studies in the context of the altmetric research, most previous approaches are essentially of an exploratory nature, studying the dissemination, reception and the conversation around academic objects in the social media. (Sugimoto et al, 2017; González-Valiente, Pacheco-Mendoza and Arencibia-Jorge, 2016).

It is thus of central importance to understand the uses and users of social media in the context of scholarly communication (Sugimoto et al, 2016). However, more contextual aspects related to the 'who', 'how', 'when' and 'where' of the reception of scientific publications in social media have been also considered to be central to altmetric research (Costas, 2017).

These more contextual aspects are still recent. Only few works have addressed this perspective, especially with regard to the qualified audience in the reception and understanding of the conversational connections (Holmberg et al., 2014; Walter, Lörcher and Brüggemann, 2019), in the analysis and identification of the public and its type of engagement (Haustein and Costas, 2015; Haustein, 2018a, 2018b; Yu et al., 2019; Joubert and Costas, 2019; Alperin, Gomez and Haustein, 2019) and the so-called 'communities of attention' (Haustein, Bowman and Costas 2015; Dıaz-Faes, Bowman and Costas, 2019).

Behavior analysis of scientists on social media is an important way of knowing not only the ways in which it is used by a scientific community, but also of who they interact with within the academy and with the public applying social media metrics based on network-based approaches. According to Wouters, Zahedi and Costas (2018) these are focused on analyzing the relationships and interactions among the different actors and these are also the least developed and more research will be necessary to fully grasp the possibilities of these analyses.

The work of Holmberg et al. (2014) sought to understand which contexts, resources and social activities influence the behavior of astrophysicists on Twitter. The study was guided by research questions that considered the influence of astrophysical activities in terms of frequency of publications, use of resources such as hashtags and connections established in the conversations maintained in the microblog. The authors found that astrophysicists communicate with a variety of types of users (e.g. colleagues, science communicators, other researchers and educators) and that in the ego networks of astrophysicist different groups of followers with diverse professional roles could be identified. The results also showed that tweets with information sharing activities were more frequent than conversations or expressing opinions.

Walter, Lörcher and Brüggemann (2019) investigated who scientists interact with on Twitter, and whether their communication differs when engaging with actors beyond the scientific community. They focus on the climate change debate on Twitter and combine network analysis with automated content analysis and the findings indicating that scientists use language strategically when communicating beyond the scientific community.

Studies of scientific tweeters will help better understand the origin and value of Twitter altmetrics (Yu et al., 2019). By asking "who is tweeting about scientific articles" the research by Haustein and Costas (2015) sought to identify groups of users who post messages about scientific articles, analyzing their self-presentations expressed in the descriptions of their profiles (i.e. Twitter "bios"), the number of followers, as well as the degree to which they engage with tweeted articles. According to the authors, analyzing the terms used in the Twitter descriptions suggests that scientific articles are tweeted by individual accounts (identifying themselves as professional, personal or both) as well as institutional accounts (organizations or interest groups). Institutional accounts have a stronger orientation towards disseminating and sharing research outputs, while academic or personal tweeters were more engaged in the discussion of the papers.

A comprehensive analysis of the scientific tweeter's productivity and geographic distribution based on a data set containing 2.63 million records collected from October 2011 to June 2016 was conducted in the study by Yu et al. (2019). The authors analyze the accounts types and identities and the different activity levels of 1468 scientific tweeters. Their results show that Scientific tweeters are widely distributed around the world but in a different pattern with the distribution of general Twitter users. In addition, scientific tweeters are found to be more active in tweeting scientific products than retweeting them in certain areas (Yu et al., 2019).

A study similar to the previous one was conducted by Joubert and Costas (2019) focusing on the understanding of the identities, characteristics and activities of South African science tweeters. Compared to non-scholarly Twitter users, the scholarly tweeters tweeted about research articles more frequently, were active on Twitter over longer periods of time, published more original tweets and used hashtags more frequently. In their Twitter bios, these scholars typically use academic terms to describe themselves, thereby presenting themselves as experts on this social media platform (Joubert and Costas, 2019).

Reflecting on how research moves between Twitter audiences Alperin, Gomez and Haustein (2019) investigate whether scientific articles shared on platforms like Twitter diffuse beyond the academic community. The case study explores a new method for answering this question by identifying 11 articles from two open access biology journals that were shared on Twitter at least 50 times and by analyzing the follower network of users who tweeted each article. The results show that diffusion patterns of scientific articles can take very different forms, even when the number of times they are tweeted is similar. For the authors the study suggests that most articles are shared within single-connected communities with limited diffusion to the public but the proposed approach and indicators can serve those interested in the public understanding of science or research evaluation to identify when research diffuses beyond insular communities (Alperin, Gomez and Haustein, 2019).

The study of Haustein, Bowman, and Costas (2015) is the first one to use the expression "community of attention". For these authors, the analysis and application of various altmetrics such as tweets for scientific works still lack adequate interpretive frameworks, mainly because the processes behind the metrics are still not fully understood. Currently, each tweet is counted equally on platforms like Altmetric.com or ImpactStory, and studies tend to ignore the type of user and tweet content, although tweets have multiple types of discussions (Haustein, Bowman, and Costas, 2015).

For Haustein, Bowman and Costas (2015) the communities of attention around scientific publications on Twitter can be further classified based on the level of engagement with the article and the exposure of the users. Engagement is measured as the degree to which the tweet text differs from the title of the tweet article. The exposure refers to the potential tweet audience, as measured by the number of followers of the user.

Finally, the study of Dıaz-Faes, Bowman and Costas (2019) are framed around the idea of developing a second generation of social media metrics, focused on characterizing the different social media communities of attention around science and their activities and interactions around scientific results. The research shows that social media metrics in science can be indicators not only of use and visibility of publications but also of interaction and dissemination of scientific knowledge across communities of attention. In addition, they can also characterize these communities (Dıaz-Faes,

Bowman and Costas, 2019). The authors draw on the overall activity of social media users on Twitter interacting with research objects and based on an exploratory and confirmatory factor analysis, four latent dimensions are identified: 'Science Engagement', 'Social Media Capital', 'Social Media Activity' and 'Science Focus' and their research breaks new ground for the systematic analysis and characterization of social media users' activity around science.

The analysis of communities of attention refers to the analysis of different communities of active users in social media platforms and their interactions with scientific outputs or entities. These communities are often defined according to the level of exposure of shared articles and they are measured according to connections between followers-followees (Wouters, Zahedi e Costas, 2018). Haustein, Bowman and Costas (2015), Dıaz-Faes, Bowman and Costas (2019) and Alperin, Gomez and Haustein (2019) claim that we can consider that the community of attention can be further qualified by the identification of users who share scientific articles and the connections established between them, by the publications they have tweeted and by their followers. Given that Twitter builds social values from users' informational and conversational practices (Recuero e Zago, 2009), additional layers of interaction – such as mentions and retweets – allows outlining not only a community of attention, but also a Community of Attention Network (CAN).

Mentions in tweets allow someone (i.e. a tweeter) to directly target a specific user (i.e. another tweeter) through the public Twitter feed or, to a lesser extent, refer to a third-person individual (Honeycutt and Herring 2009). Retweets are different, they act as a form of dissemination, allowing individuals to relay content generated by other users, thereby increasing the visibility of content (boyd, Golder and Lotan 2010).

The exposure, as posed by Haustein, Bowman and Costas (2015) and Alperin, Gomez and Haustein, 2019), indicates, from a structural aspect of Twitter, the potential audience that a tweet can achieve, given the number of followers of the user. In this context we also add two other structural aspects: (1) the direct and specific indication of what is shared when mentioning another user, in such case, we may be talking about a degree of recommendation; and (2) the spreading of content originally shared by another user, we may consider it to be a degree of approval or endorsement.

Altmetrics have been applied to measure online attention (engagement or influence) and societal impact among different audiences, which can be used to map interactions, contexts, networks and communities (Holmberg, 2017; Dıaz-Faes, Bowman and Costas, 2019). To better understand the communities of attention on Twitter is necessary to consider the different types of interactions that users have in the microblogging platform, and what they mean.

In addition to broadcasting one's message via tweets, Twitter users can also directly address users with @mentions or @replies on an interpersonal level (Haustein, 2018a). 'Mentions' (@user messages) are used to reference other users. They are considered a form of "addressivity" (Honeycutt and Herring, 2009) or 'attention-seeking', in order to gain the target user's attention for a specific message, which is essential for a conversation to occur (boyd, Golder and Lotan, 2010), alerting the mentioned user that they are being talked about (boyd, Golder and Lotan, 2010).

One of the most important features of Twitter is the support for "retweets" or messages re-posted verbatim by a user that were originated by someone else. Retweeting indicates not only interest in a message, but also trust in the message and the originator, and agreement with the message contents (Metaxas et al, 2014; McNeill and Briggs, 2014; Metaxas et al, 2015). Retweets represent a specific form of information diffusion and seem to play a significant role in sharing scientific papers on Twitter Alperin, Gomez and Haustein (2019). As sharing information is one of the main motivations for scholarly Twitter use, retweeting is likely to be common among tweeting academics (Haustein, 2018a) and also serves as social purposes such as public endorsement and exhibition of support (boyd, Golder and Lotan, 2010; Kim and Yoo, 2012) even though they are not considered that way by journalists (Haustein, 2018a).

From an operational point of view, the "addressivity" or recommendation level can be measured by the percentage of tweets with mentions to other users in a given set of tweets. Somehow, mentions of other users can be considered as a form of alertor a way of hinting the mentioned user to read the tweet. Retweets can be seen as a form of dissemination by endorsement or exhibition, thus the percentage of retweets in a set of tweets can be considered as an indication of content approval action.

These two Twitter affordances - mentions and retweets - serve different and complementary purposes. Together they act as the primary mechanisms for explicit and public interaction between users and users on Twitter (Conover, 2011) and should be considered by altmetric studies to study the formation of "communities of attention networks".

## Material and methods

This paper presents an exploratory descriptive qualitative study case aiming at understanding the community of attention network formed around a scientific paper. Taking into consideration that studies on publications mentioned on social media indicate that altmetric values vary between areas, journals and subjects (Costas, Zahedi and Wouters, 2015; Alperin, 2015) we chose to study the altmetric data analysis of a single article.

We start by reusing the data collected by Araújo, Oliveira and Lucas (2017) who analyzed the 10 most popular articles in terms of online attention with data from Altmetric.com, present in the SciELO Brazil Collection of the Science Open directory. To exemplify the potential of forming a community of attention network was necessary to consider an article with high numbers of retweets and mentions. Additional reasons for choosing this article were:

- is published by a Brazilian academic journal in Portuguese language;
- is well positioned as to the attention score in Altmetric.com;
- has an index of conversational tweets (CT) greater than the information tweets (IT). Thus, CT = tweets with mentions and retweets degree [(mentions+retweets)/total tweets]; IT = regular tweet degree [tweets without mentions or retweets/total tweets]; where CT > IT.

The article that best met the criteria was written by Marcio C. Mancini and Alfredo Halpern published in 2002 in the journal Arq Bras Endocrinol Metab with 736 total tweets, 210 mentions, 495 retweets and 31 regular tweets (CT = 0.957 and IT = 0.042). The paper is an extensive review on the criteria for evaluating anti-obesity treatments and on pharmacological derivatives drugs and presented an analysis of all clinical studies of more than ten weeks duration with drugs used in the treatment of obesity (Mancini and Halpern, 2002).

Twitter user data such as number of followers representing exposure and mentions to other profiles representing the recommendation were collected from Altmetric.com. The description information for each profile was obtained by extracting information bio's accounts from Twitter on September 27, 2017.

The article's altmetric data from Twitter were analyzed using Conversation Analysis (CA) techniques based on MOOD (Microanalysis Of Online Data perspective), an appropriate and a qualitative method for analyzing elements of online interaction to the type of rigorous conversation analysis in digital environments such as social media (Giles et al., 2015). This approach uses techniques for online analysis and enables to develop tailored analysis modes for specific forms of online data (e.g. 'tweets' on Twitter).

The CA served to characterize users who tweet, retweet and those who are mentioned in tweets, and to understand the timeline of tweets in their contextual aspects. The user's information profile was analyzed using clusters of the terms, a visualization technique by VOSViewer. In addition, the software Gephi was used to generate a visualization of the community of attention network through the analysis of the "addressivity" or recommendations (by mentions) and spreadings or endorsement (retweets).

VOSviewer uses a part-of-speech tagging algorithm that is only optimized for English grammar and English terms. Since the Twitter bios and extracted terms in this study are mainly from Brazil and thus written in Portuguese, each user profile information was submitted to a required group of pre-processing operations following the steps adopted by Pereira et al (2017) below: (a) Lowercasing: Every message presented in a tweet was converted into lowercase; (b) Cleaning Entities: Removing URLs, user mentions, hashtags and digits from the text message; (c) Lemmatization: Transformation of plural words into singular ones; (d) Punctuation Removal: Every punctuation was removed as well as smiles or even emojis; (e) Stop Words Removal: The removing of this kind of words was made using the Portuguese NLTK dictionary; (f) Short Tokens Removal: Words such as 'kkk', 'aaa', 'aff' and other of the same style were removed.

**Results**

Altmetric.com's general data with all online attention details for the article <http://www.altmetric.com/details/1844847> indicated a total of 736 tweets from 134 users with a combined exposure greater than 459,018 followers. The Twitter life span of the paper is presented in Figure 1. This type of analysis reveals the amount of time between the first and last tweet, indicating how long a document stays relevant on Twitter (Haustein, 2018).

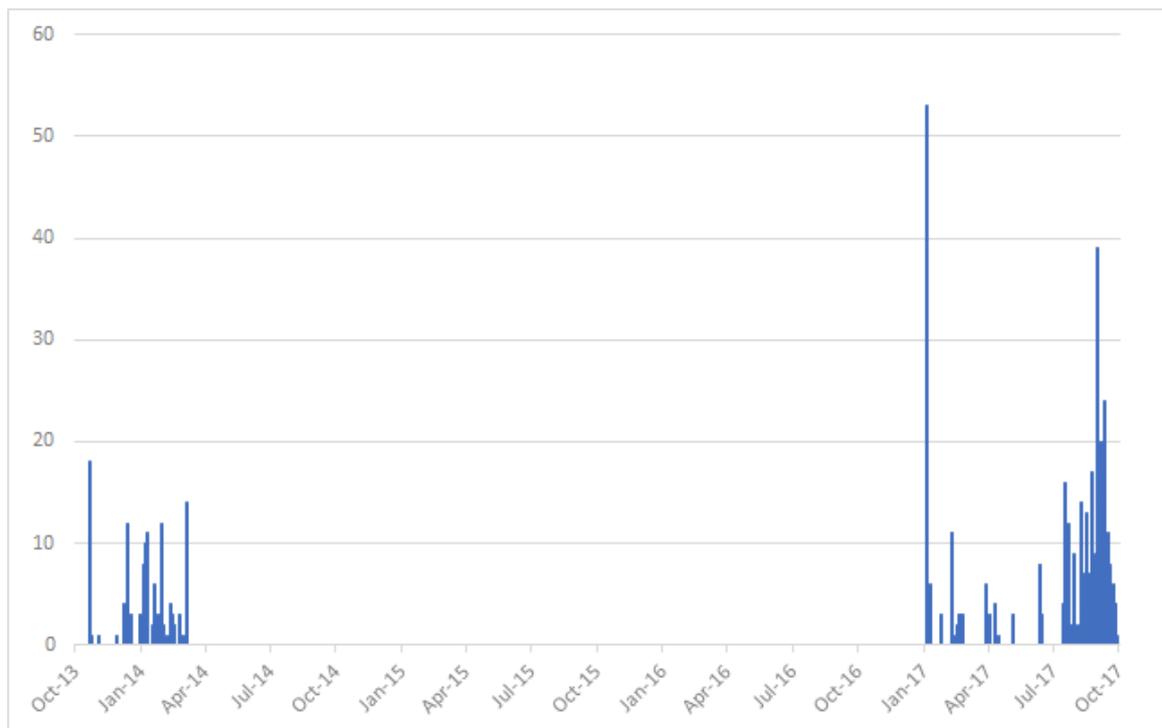

Figure 1. Twitter life span of the paper over period analyzed

The response time and Twitter life span were found to be contrary to what is generally common in high-indexing and rapid-dissemination (Priem et al, 2010) with article audience peaking shortly after publication and significant decline after the first few weeks, or even in the early days (Haustein, 2018).

As for the profile of users based on Altmetric data, 96% are members of the public, only 1% of health professionals and scientists, and 2% are unknown. A challenge to qualify online attention in identifying "who" shares scientific articles and "who" recommends is that not every user describes basic information in their profile, which compromises his characterization. 52.8% of those who share the article had this information in their profile; a term cluster is shown in Figure 2. As for the users mentioned in shared tweets, 89% had information in "Twitter's bio"; a term cluster is shown in the Figure 3.

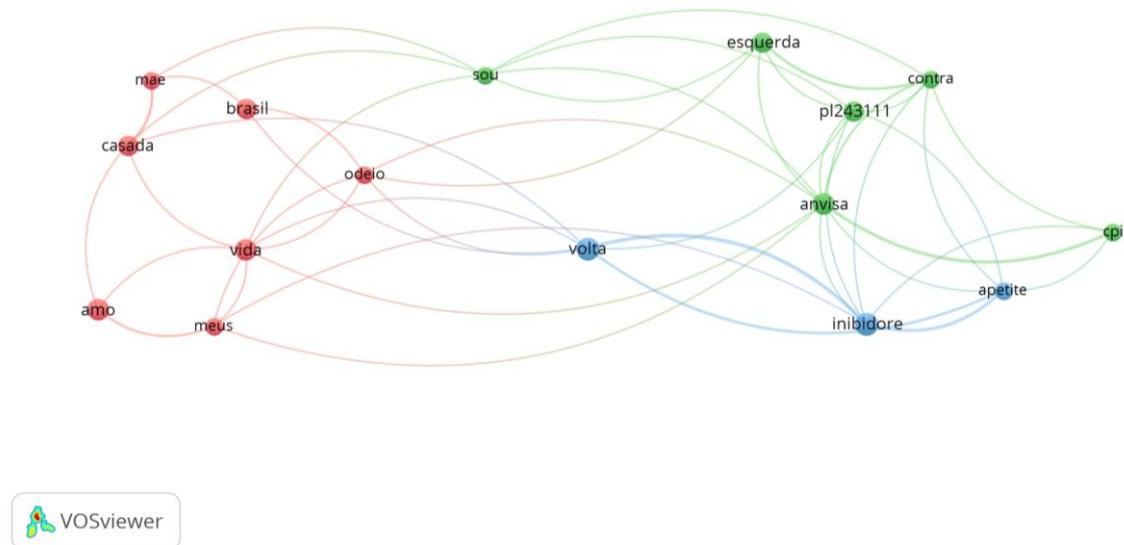

Figure 2. Cluster of self-description terms in the profile of those who share the paper

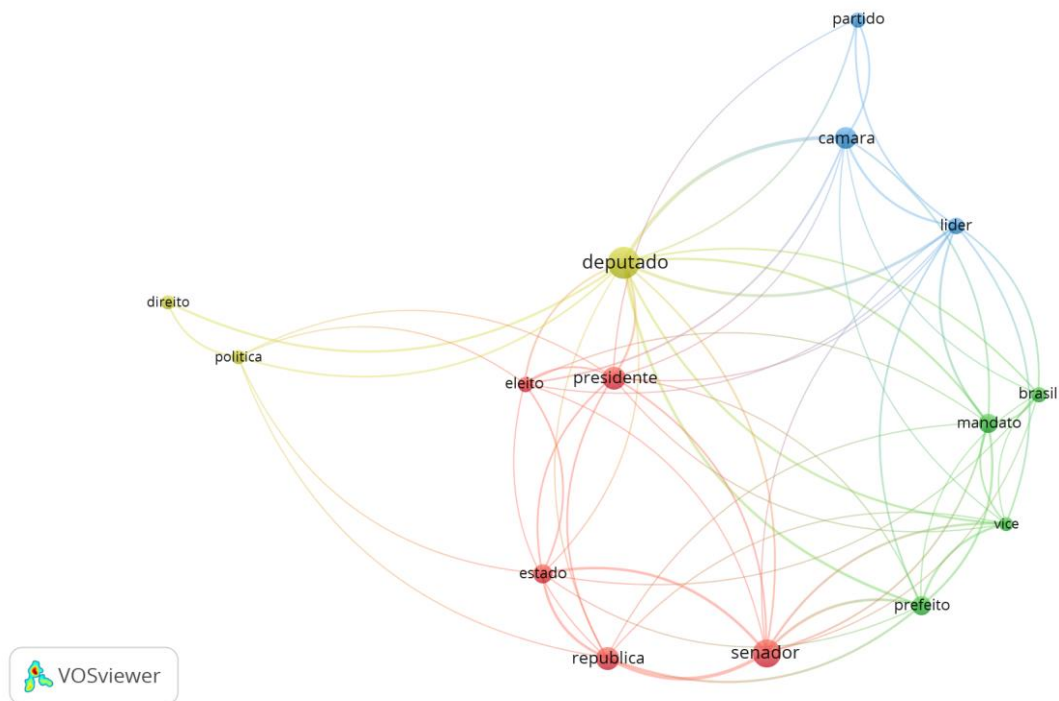

Figure 3. Cluster of self-description terms in the profile of who is mentioned

According to Marwick and boyd (2010) people tend to present themselves in fixed, singular, and self-conscious ways and the microblogging site Twitter affords dynamic, interactive identity presentation to unknown audiences. The figures show us quite different profiles depending on whether the users share the paper or they are mentioned together with the paper.

In addition to the profile description, self-presentation on Twitter also takes place through ongoing 'tweets' and interactions with others, rather than only the static profiles (Marwick and boyd, 2010). On analyzing the tweet contents, we noted that online attention around the paper is related to the strategic use of the paper as a resource of online activism, and as an instrument of digital advocacy. This aligns with the role of social media like Twitter for the mobilization of interested parties for activities of consciousness-raising and support (Guo and Saxton, 2014).

The timeline of received online attention in Table I, despite being sparse, confirms this is illustrated in three distinct moments, which are discussed below. The digital advocacy practiced by users who shared the paper indicates different stages of engagement and mobilization. All examples listed in the Table have been translated from Portuguese to English and the original tweet can be checked on the status link to the corresponding status link.

Table I. Examples of "mobilization" and digital advocacy tweets

| Moments | Examples of tweets |
|---|---|
| First group of tweets | Example 1: "@user With a medical prescription these drugs are great, studies that prove it http://t.co/OlpLAmIl3f" 20 Oct 2013 from: https://twitter.com/politica_estilo/status/391900057881939969<br><br>Example 2: "@user position ABESO http://t.co/30cy2PsrLj … … Position ABRAN http://t.co/fZ3BDHw672 … http://t.co/HbEmcNkFZF" 12 Dec. 2013 from: https://twitter.com/anton_zack/status/411155591126478848 |
| Second group of tweets | Example 3: "@parliamentary_user #PL2431_11 medication was 1of the pillars and there are no miracles http://t.co/XXy2v0BPHk http://t.co/uBElzMWAj1 http://t.co/i4ShuPlNhp" 14 Feb 2014 from: https://twitter.com/sacred_killer/status/441348533380079616<br><br>Example 4: "@parliamentary_user they made a mistake http://t.co/XXy2v0BPHk http://t.co/58ESdmzbIm http://t.co/2zudRbVRiY http://t.co/45vcONe2lt http://t.co/i4ShuPlNhp" 5 Mar 2014 from: https://twitter.com/sacred_killer/status/441350419499855872 |
| Last group of tweets | Example 5: "@user https://t.co/mJ2gqzVU4K we need help, Anvisa has the studies" 5 Jan 2017 from: https://twitter.com/soniajcalixto/statuses/816969353396686848<br><br>Example 6: "https://t.co/7qHqGBYIg3 @parliamentary_user how do they say there are no studies? The proof that they exist is here" 5 Jan 2017 from: https://twitter.com/soniajcalixto/statuses/816973687387787264<br><br>Example 7: "@parliamentary_user Good morning deputy, the Anvisa said that there are no studies on inhibitors but we do have https://t.co/3AsEBHZqjk on the agenda of Chamber of Deputies #pl2431_11" 05 Jan 2017 from: https://twitter.com/soniajcalixto/statuses/816980071739256832<br><br>Example 8: "@user #LEI_13454 To read → → Pharmacological Treatment of Obesity https://t.co/LrkwOhiVi9 ← ← #PL2431_11 to read. https://t.co/qJjQOIGMns 20 Jul 2017 from: https://twitter.com/domlusa1/statuses/888174907640741888<br><br>Example 9: @parliamentary_user the project took 7 years. It is now closed Pharmacological Treatment of Obesity http://www.scielo.br/scielo.php?scr... #PL2431_11 #LEI_13454_1 set 2017 from: https://twitter.com/domlusa1/status/903625960192450560 |

The hashtags also reinforce the public's understanding and beliefs about drug effects and even the regulation of its use. In the case study, even, they indicate different moments, one for the approval of Law Project nº 2431/2011 (#PL2431_11) and the other in celebration of a lawsuit attended nº 13454/2017 (# LEI_13454). According to Van Honk and Costas (2016) the hashtags work as a linking element used to create impromptu categorizations of tweets but also the means for Twitter users to follow the "conversation" on a topic and communicate with similar communities of interest around the hashtag. Thus, hashtags are instruments to enlarge the potential audience and can be seen as an element related to the higher reception of scientific publications (Van Honk and Costas, 2016) expanding the potential exposure of their tweets to other users beyond their original set of followers (Wouters, Zahedi and Costas, 2018).

In the general data we have 736 tweets, which present:

- the degree of sharing (informative tweets): 31 (4.21%)
- Recommendation level: 210 mentions (28.53%)
- the degree of spreading (RT): 495 retweets (67.26%)

The community of attention network was mapped from these interactions, which can be seen in Figure 4. This is a behavioral network with unidirectional or bidirectional interactions with 242 actors, 571 connections and 27 disconnected nodes (actors without interaction). We used the distribution of ForceAtlas 2 with behavior alternative option to dissuade hubs and avoid overlap with the parameters reconfigured gravity and approximation to 1.0 and 1.2, respectively.

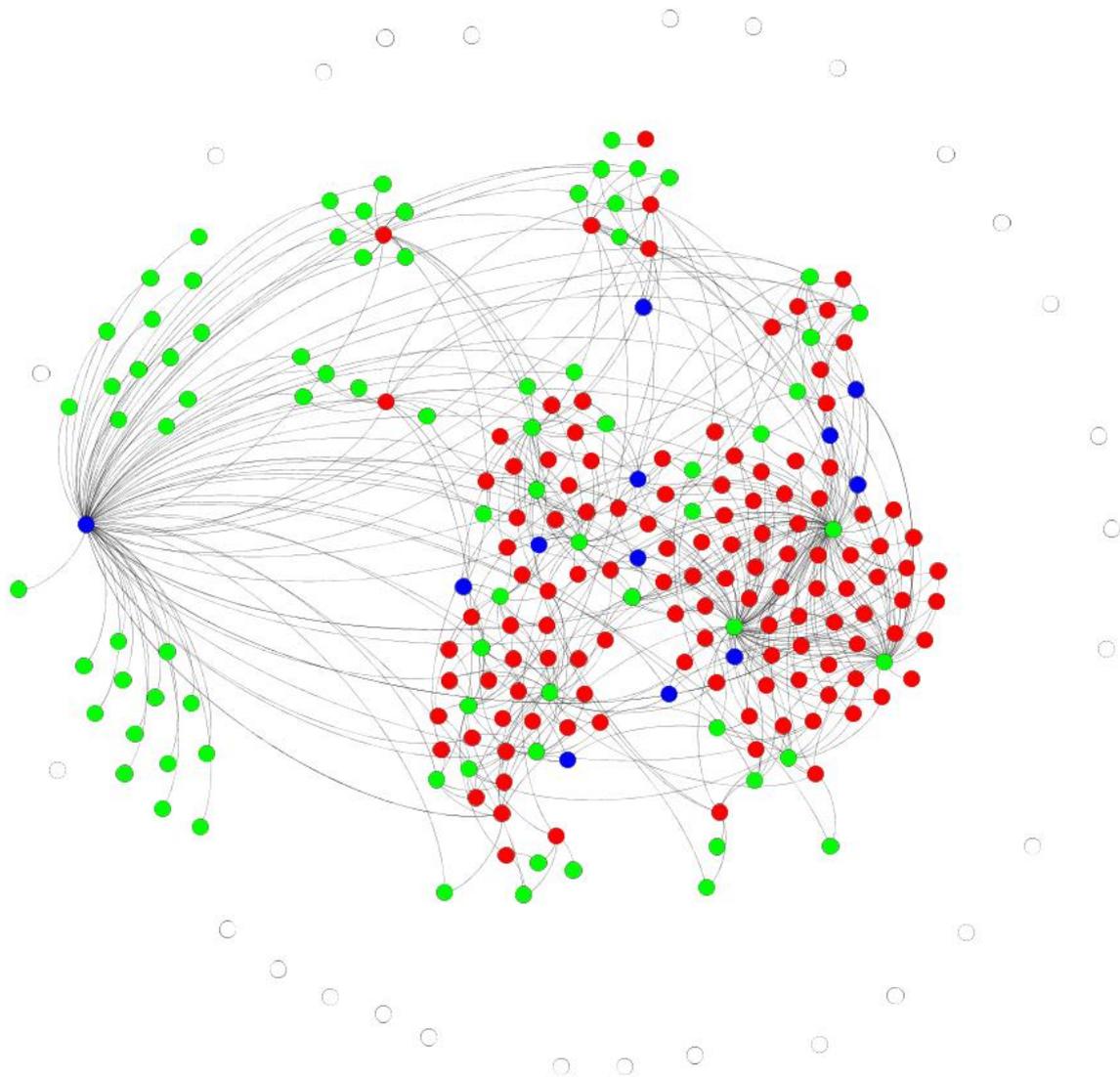

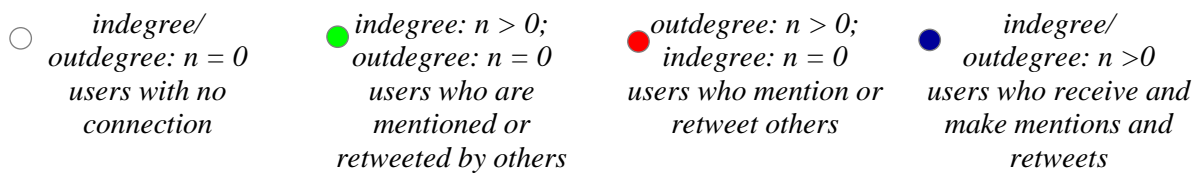

Figure 4. Communities of attention network around the paper by mentions and retweets

The white nodes represent the users with informative tweets; those in green color only have value of indegree, that is, they are mentioned or retweeted. Nodes of red color only have value of indegree, that is, they mention or retweet, and finally, the blue ones play both roles in the network, either mention or retweet, as they are mentioned and / or retweeted. The scatter plot and logarithmic scale of Figure 5 assists in this representation.

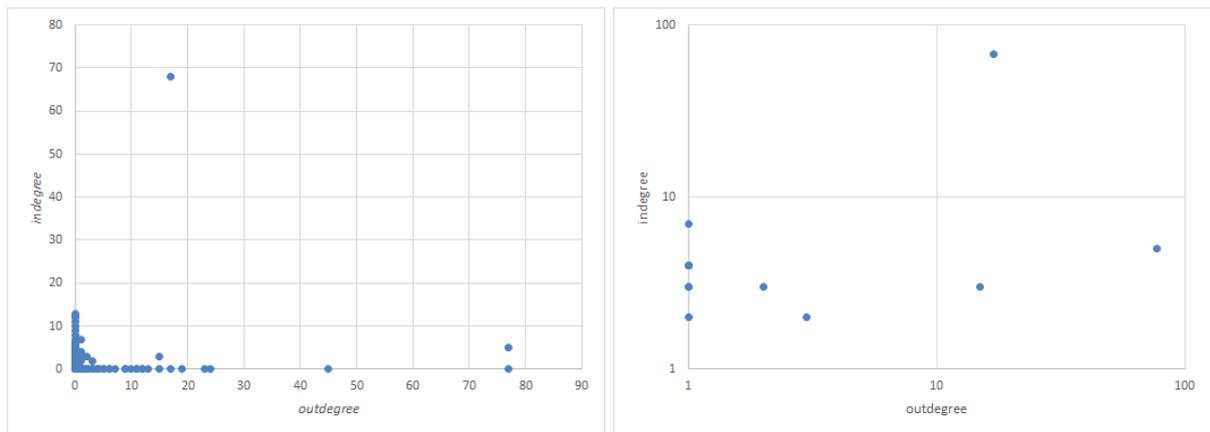

Figure 5. Scatter plot and logarithmic scale chart

More than 80% of the actors present only indegree or outdegree values being well connected to the corresponding axis, being 122 users with only values for outdegree (50%), such as nodes 35 (outdegree = 12) and 61 (outdegree = 13). And 81 users with only indegree level performance (33%), with highlight for node 167 (indegree = 77) and node 205 (indegree = 45). Both nodes are news media-related accounts. We can observe that of all the actors with indegree value = n> 1 (54 users), only 4 have some outdegree, such as node 157 (indegree = 77; outdegree = 5); and that of all actors with outdegree value = n> 1 (131 users), only 12 have indegree value, with highlight for node 17 (outdegree = 68 and indegree = 17).

**Discussion**

The paper analyzed was published in 2002 and even we know Twitter was created in 2006 and Altmetric started tracking attention to research across various attention sources in October 2011 the first recorded tweet was in October 2013.

In this case, we can argue that the response time of this paper on Twitter is in accordance with what is referred to in the literature on the "sleeping beauty" or sleep beauty effect (van Raan, 2004; Ke et al, 2014). The concept of "sleeping beauties" is used in citation studies to measure the period between the date of publication of a document and the first citation it receives, being publications labeled as "sleeping beauties" when they have a particularly long citation delay. For Haustein (2018) in the context of altmetrics, the sleep beauty effect it would be equivalent to the delay between the date of publication and the appearance of the first tweet (or any other social media event).

The "exposure time" could be defined as the lifetime of the article on Twitter. It measures the number of days between the first and last tweet and indicates how long a document remains relevant on the network (Haustein, 2018). The article recorded a floating exposure time as illustrated in Figure 1. Although it takes five years, from 2013 to 2017, there were no activities for the years 2015 and 2016, when it "fell asleep" again.

Profile descriptions are also an appealing source for outlining the potential communities of users by providing hints on the background and professional activity of users (Dıaz-Faes, Bowman and Costas, 2019). The analysis of the profile of users shows in the first case (Figure 2) a higher incidence of individual profiles with self-presentation focused on personal issues (red cluster on the left), with terms such as "casada" (married), "meus" (mine), "amo" (I love), "mãe" (mum), "vida" (life) which reveal tweeters who disclose their private and personal life (Dıaz-Faes, Bowman and Costas, 2019).

Marwick and boyd (2010) argue that symbolic interactionism claims that identity and self are constituted through constant interactions with others, thus self-presentation of users is constructed in a collaborative fashion. For these authors "individuals work together to uphold preferred self-images of themselves and their conversation partners, through strategies like maintaining (or 'saving') face, collectively encouraging social norms, or negotiating power differentials and disagreements". Figure 2 profiles also reveal strong civic and political engagement and a declared predilection mostly for the political right (green cluster on the top-right) with descriptions such as "sou contra a esquerda" (I'm against left-wing parties) or "odeio a esquerda" (I hate left-wing parties), "cpi" (acronym for "Comissão Parlamentar de Inquérito" that means Parliamentary Commission of Inquiry), and a militancy fighting obesity (blue cluster on the bottom-right) with an emphasis on the expression "quero inibidores de apetite de volta" (I want appetite inhibitors back).

In the second case (as shown in Figure 3), the analysis of the frequent terms of tweeters mentioned alongside the paper indicates personal-professional and institutional profiles of the national political scene, together with descriptions of their areas and activities. These are of political and parliamentary actors, such as "deputado" (federal deputy) - (yellow cluster); "senador da república" (senator) and "presidente do partido" (political party leader) - (red cluster on bottom-center); "líder da camara federal" (leader of the chamber of deputies) and "partido" (political parties) – (blue cluster on the top-right); and local authority as "prefeito" and "vice-prefeito" (mayor and deputy mayor) - (green cluster on the bottom-right).

To better understand how the paper spread across the network within and between sub-communities it is necessary to look at the kind of sharing that took place (i.e. original tweet or retweet) and at the timeline when the article was tweeted (Alperin, Gomez and Haustein, 2019). Table I shows three groups that indicate different moments of digital advocacy practices identified by the Twitter users who shared the paper. The first group of tweets (from 2013) in which the users use the paper for credibility and consciousness-raising regarding the efficacy of appetite inhibitor (Example 1) and emphasize the positioning of associations/specialist societies on the topic, like ABESO – Brazilian Association for the Study of Obesity and Metabolic Syndrome, and ABRAN – Brazilian Association of Nutrology (Example 2).

There is a second group of tweets (from 2014), in which the users engage in "popular pressure" to the parliamentary representatives by asking for support and a vote in favour of passing the Law Project nº 2431/2011. This law prohibits ANVISA[1] from vetoing the production and commercialization of the anorexigenics: sibutramine, amfepramone, femproporex and mazindol (BRASIL, 2017). There are also other external links to Public Audiences (Example 3) and decisions of the judiciary power against the contents of the Bill (Example 4).

As a consequence of the non-approval of the Law Project for more than three consecutive sessions in the plenary session of the Chamber of Deputies in December 2016; the last tweets (from 2017), remind parliamentarians that there are studies on appetite inhibitors and that its discussion and vote are still kept on the agenda (examples 5, 6 and 7); and continues after the Law Project nº 2431/2011 has become a Law nº 13454/2017 (example 8 and 9), which authorizes the production, commercialization and consumption, with a medical prescription, of anorexigenics: sibutramine, amfepramone, femproporex and mazindol (BRASIL, 2017).

When analyzing the usefulness of altmetrics for pharmacoepidemiologists Gamble et al (2017) emphasize that it may be used to evaluate the public's understanding, knowledge, attitude, and beliefs about drug effects. The analysis of the table with regard to the types of tweets that circulate about the article allows us to confirm that there is an activity centered on online activism by the users who share it in the form of actions of awareness and support.

Informative tweets that shared the paper without any conversational aspect to another user in the network have less representation (4.21%). Conversational tweets (mentions and retweets together) exceed and reach 95.79%. The recommendations of the article with tweets made through mentions to other users represent 28.53%.

Mentions on Twitter are the best resources to start a conversation, whether to direct a tweet to a specific recipient or to refer to another user (Honeycutt and Herring 2009). Spreading of messages in retweets represent 67.26% of all tweeting activity, similarly to the results of Alperin, Gomez and Haustein (2019) more than half of all tweets for the paper were retweets. Retweeting can be understood as both a form of information diffusion and participation in a diffused conversation. Thus, we may consider that spreading tweets is not simply sending messages to new audiences, but also to validating and engaging others (boyd, Golder and Lotan 2010).

The Communities of attention network illustrated in Figure 4 shows a network of nodes, with a mean of 2.36 - indicating that (on average) each user is connected with 2 other users, that is, each actor interacted with just over 2 actors - on average. The diameter of the graph, which calculates the maximum distance between the users is of 6.0, that is, one actor is distant of another size in the maximum 5 intermediaries. The density of the graph is 0.02 in a network in which the approaches are well together. This means that of all potential network access, all work interacting with all types of data outputs, this practice 0.2%.

We observed that few nodes concentrate most of the relationships, in a position of centrality in the network. For the most part, the relations established between the nodes are uni-directed, so that there are well-established roles among those who share or refer to a user (outdegree) and those who are mentioned or retweeted (indegree). As in Alperin, Gomez and Haustein (2019) only a small minority of users were isolated and had no connection to the core group.

The Figure 5 by the scatter plot and the logarithmic scale that clearly describe details at low concentration area, assists in this representation and these values confirm that in this community of attention network the users adopt clear and well-defined roles, since whoever is mentioned or retweeted usually does not mention or retweet others. From the number of users analyzed in the Community of Attention Network, the node 17 can be considered as the most activist engaged in the cause of the fight against obesity. She is one of the most active users tweeting mentions to most parliamentarians and being retweeted. After the paper was "asleep" in 2015 and 2016, she posted the first "wake up" tweet of online attention in 2017, accounting for 64% of the tweets of the first week of January 2017, exactly the peak period of Twitter life span of the paper as shown in Figure 1.

**Conclusion**

In altmetric research - given the dynamic, fluid and momentary aspects of most of the informational devices of the social web, among which social media stand out - the methodological designs to study the behavior of mentions and social media interactions vary enormously, making the experiments of difficult spreading. But we emphasize that behavioral differences among social media actors should not only be viewed as an important element necessary to qualify the altmetric studies.

These contextual and behavioral aspects must be perceived as fundamental elements for the better characterization of users and interactions in the social media landscape.

In this study, we have studied the online attention of one paper. We have qualified its online attention by using the Conversation Analysis (CA) techniques based on MOOD. Thus, we have studied not only who shares the article, but also who is mentioned directly in the tweets, together with the analysis of the retweets. The community network of attention of the analyzed paper presented actors with well-defined roles as the actions of informative tweets, mentions or retweets.

Regarding the values of the graph of the network formed, since it is the first study that considers a community of attention network, which takes into account the interactions around shared articles, it is still not possible to say if they are high or low and other studies are required from this perspective. The density, for example, in absolute terms is a low value, but requires future parameters to measure whether this is a comparatively low, medium or high value with other investigations.

The combined contextual approach presented in this paper (combining actors, mentions, and retweets) introduces a contextual perspective that analyzes with more precision who are the actors that interact. This approach also offers an additional visual and a quantitative sense of the social engagement around the article. For example, in our study it was possible to prove that this article was used for the purposes of digital advocacy, in which individuals used scientific knowledge to raise awareness and 'pressure' parliamentarians and specialized societies on their lawsuit (specifically the approval of the Law Project 2431/2011 in Brazil). Although initial studies suggest that social media has rather opened a new channel for informal discussions among researchers, rather than a bridge between the research community and society at large (Sugimoto et al, 2017) this case study can prove that this does not always happen.

The study of how social media users interact with research objects will help to pave the way to further unravel the mechanisms by which academic and, especially, non-academic actors interplay with scientific outputs and scholarly entities (Dıaz-Faes, Bowman and Costas, 2019). As such, the combined and contextual approach introduced in this study can be seen as a stepping-stone towards more advanced social media metrics or better social media studies of science as promoted by Costas (2017), which could be seen as the study of the relationships and interactions between social media and scholarly objects. Thus, research wouldn't just be circumscribed to the study of the reception of scholarly objects in social media (the predominant approach of most altmetric studies), but also to how scholarly entities interact with other social media actors (Costas, 2017) paving the path to a better understanding of the social impact of altmetrics an aspect that indeed still requires more future research.

**Acknowledgements**

The author would like to thank the National Council for Scientific and Technological Development - CNPq (426777/2016-6) and the Alagoas Research Foundation - FAPEAL (600301057/2016) for the financial support. The author would also like to thank Fábio Gouveia for comments, Rodrigo Costas for proofreading and English review and Gustavo Caran for his help in analyzing the graph data.

**References**

Alperin, J. P., Gomez, C. J, & Haustein, S. (2019). Identifying diffusion patterns of research articles on Twitter: A case study of online engagement with open access articles. *Public Understanding of Science*, 28(1),2–18. DOI: https://doi.org/10.1177/0963662518761733

Araujo, R. F., Oliveira, M. & Lucas, E. R. O. (2017). Altmetria de artigos de periódicos brasileiros de acesso aberto na ScienceOpen uma análise das razões de menções. *Reciis – Rev Eletron Comun Inf Inov Saúde*. 11 (sup), 1–7. https://www.reciis.icict.fiocruz.br/index.php/reciis/article/view/1376

Araujo, R. F. & Furnival, A. C. M. (2016) Comunicação científica e atenção online: em busca de colégios virtuais que sustentam métricas alternativas. *Informação & Informação*, 21(2), 68–89. DOI: http://dx.doi.org/10.5433/1981-8920.2016v21n2p68

boyd, D., Golder, S. & Lotan, G. (2010). Tweet, tweet, retweet: Conversational aspects of retweeting on twitter. Proc. Hawaii Intl. Conference on Systems Sciences, (43), 1–10. *Proceedings…* Honolulu, HI, USA. DOI: http://dx.doi.org/10.1109/HICSS.2010.412

Brasil, Câmara dos Deputados. *Projetos de lei e outras proposições: PL 2431/2011*, Retrieved from: http://www.camara.gov.br/proposicoesWeb/fichadetramitacao?idProposicao=522126

CGI. Comitê Gestor da Internet no Brasil (2017). *TIC domicílios 2016*: pesquisa sobre o uso das tecnologias de informação e comunicação nos domicílios brasileiros. Núcleo de Informação e Coordenação do Ponto BR: São Paulo, pp.430

Costas, R. (2017) Towards the social media studies of science: social media metrics, present and future. *Bib. An. Invest.* 13(1), 1–5. http://revistas.bnjm.cu/index.php/anales/article/view/4192

Costas, R., Zahedi, Z., & Wouters, P. (2015). The thematic orientation of publications mentioned on social media: Large-scale disciplinary comparison of social media metrics with citations. *Aslib Journal of Information Management*, 67(3), 260–288. DOI: http://doi.org/10.1108/AJIM-12-2014-0173

Conover, M. D., Ratkiewicz, J., Francisco, M., Goncalves, B., Flammini, A. & Menczer F. (2011, July). Political Polarization on Twitter. Presented at the 5th International AAAI Conference on Weblogs and Social Media, ICWSM: Barcelona, Spain. https://www.aaai.org/ocs/index.php/ICWSM/ICWSM11/paper/view/2847

Dıaz-Faes, A. A, Bowman, T. D., & Costas, R (2019) Towards a second generation of 'social media metrics': Characterizing Twitter communities of attention around science. *PLoS ONE*, 14(5), e0216408. DOI: https://doi.org/10.1371/journal.pone.0216408

Gamble, J. M., Traynor, R. L., Gruzd, A., Mai, P., Dormuth, C. R. & Sketris, I. S. (2018). Measuring the impact of pharmacoepidemiologic research using altmetrics: a case study of a CNODES drug-safety article. *Pharmacoepidemiol Drug Saf*. 1–10. DOI: https://doi.org/10.1002/pds.4401

Giles, D., Stommel, W., Paulus, T., Lester, J. & Reed, D. (2015) Microanalysis Of Online Data: The methodological development of "digital CA", *Discourse, Context & Media*, 7, 45–51. DOI: https://doi.org/10.1016/j.dcm.2014.12.002

González-Valiente, C. L., Pacheco-Mendoza, J. & Arencibia-Jorge, R. (2016), A review of altmetrics as an emerging discipline for research evaluation. *Learned Publishing*, 29: 229–238. DOI: https://doi.org/doi:10.1002/leap.1043

Guo, C. & Saxton, G. D. (2014). Tweeting social change: How social media are changing nonprofit advocacy. *Nonprofit Voluntary Sector Quarterly*, 43(1), 57–79. DOI: http://dx.doi.org/10.1177/0899764012471585

Haustein, S. (2018a). Scholarly Twitter metrics. In W. Glänzel, H.F. Moed, U. Schmoch, & M. Thelwall (Eds.), *Handbook of Quantitative Science and Technology Research*. Springer https://arxiv.org/abs/1806.02201

Haustein, S. (2018b). Never put off till tomorrow, what you can tweet today or: how quickly research papers spread on twitter. *Altmetric Blog*, 5, jul. https://www.altmetric.com/blog/never-put-off-till-tomorrow-what-you-can-tweet-today-or-how-quickly-research-papers-spread-on-twitter/

Haustein, S., & Costas, R. (2015, November) *Identifying Twitter audiences:* who is tweeting about scientific papers? Presented at the ASIS&T SIG/MET Metrics 2015 workshop. Retrieved from https://www.asist.org/SIG/SIGMET/wp-content/uploads/2015/10/sigmet2015_paper_11.pdf

Haustein, S., Bowman, T. D., & Costas, R. (2015). *"Communities of attention" around scientific publications: who is tweeting about scientific papers?* Presented at the Social Media & Society 2015 International Conference, Toronto, Canada. Retrieved from https://www.slideshare.net/StefanieHaustein/communities-of-attention-around-journal-papers-who-is-tweeting-about-scientific-publications

Haustein S, Peters I, Sugimoto CR, Thelwall M, Larivière V. (2016) Tweeting Biomedicine: An Analysis of Tweets and Citations in the Biomedical Literature. *Journal of the Association for Information Science and Technology*. 65(4), 656–669. DOI: https://doi.org/10.1002/asi.23101

Honeycutt, C. & Herring, S. C. (2009, January). *Beyond microblogging: Conversation and collaboration via Twitter.* Paper presented at the System Sciences, HICSS '09. 42nd Hawaii International Conference, Big Island, HI, USA. https://doi.org/10.1109/HICSS.2009.89

Holmberg K, Bowman TD, Haustein S, Peters I (2014) Astrophysicists' Conversational Connections on Twitter. *PLoS ONE*, 9(8): e106086. DOI: https://doi.org/10.1371/journal.pone.0106086

Joubert, M., & Costas, R. (2019). Getting to Know Science Tweeters: A Pilot Analysis of South African Twitter Users Tweeting about Research Articles. *Journal of Altmetrics*, 2(1): 2. DOI: https://doi.org/10.29024/joa.8

Ke, Q., Ferrara, E., Radicchi, F. & Flammini, A. (2015). Defining and identifying Sleeping Beauties in science. *Proc Natl Acad Sci,* 112: 7426–7431. DOI: www.pnas.org/cgi/doi/10.1073/pnas.1424329112

Kim, J. & Yoo, J. (2012, June). *Role of sentiment in message propagation: Reply vs. retweet behavior in political communication*. Paper presented at the International Conference on Social Informatics, Lausanne, Switzerland, 131–136. DOI: https://dx.doi.org/10.1109/SocialInformatics.2012.33

Mancini, Marcio C., & Halpern, Alfredo. (2002). Tratamento Farmacológico da Obesidade. *Arquivos Brasileiros de Endocrinologia & Metabologia*, 46(5), 497–512. DOI: https://dx.doi.org/10.1590/S0004-27302002000500003

McNeill, Andrew & Briggs, Pamela (2014, April) *Understanding Twitter Influence in the Health Domain: a social-psychological contribution*. Paper presented at the 23rd International Conference on World Wide Web, Seoul, Korea, 673–678. DOI: http://dx.doi.org/10.1145/2567948.2579280

Metaxas, P., Mustafaraj, E., Wong, K., Zeng, L., O'Keefe, M., & Finn, S. (2014, November). *Do retweets indicate interest, trust, agreement?* Paper presented at the Computation and Journalism Symposium, New York, NY. 1–5 https://arxiv.org/abs/1411.3555

Metaxas, P., Mustafaraj, E., Wong, K., Zeng, L., O'Keefe, M., & Finn, S. (2015, May). *What Do Retweets Indicate? Results from User Survey and Meta-Review of Research*. Paper presented at the 9th International AAAI Conference on Web and Social Media, Oxford, England, 658–661

Marwick, A., & boyd, D. (2010). I tweet honestly, I tweet passionately: Twitter users, context collapse, and the imagined audience. *New Media and Society*, 13(1), 96–113. DOI: https://dx.doi.org/10.1177/1461444810365313

Nelhans, G.; Lorentzen, D.G. (2016) Twitter conversation patterns related to research papers. *Information Research*, 21(2), http://InformationR.net/ir/21-2/SM2.html   acesso em 14 set.

Olijhoek, T. (2011). *Open Access Week 2011*: A Short History of Open Access. Openaccessweek Blogs. oct., 27. Retrieved from https://goo.gl/917czi

Pereira, J., Pasquali, A., Saleiro, P., Rossetti, R. & Cacho, N. Characterizing geolocated tweets in brazilian megacities. In 2017 International Smart Cities Conference (ISC2), pages 1–6, Sept 2017. DOI: https://doi.org/10.1109/ISC2.2017.8090832

Priem, J., Taraborelli, D., Groth, P. & Neylon, C.  (2010). *Altmetrics:* A manifesto, 26 October 2010. Retrieved from http://altmetrics.org/manifesto

Recuero, R. & Zago, G. (2009). *Em busca das "redes que importam":* redes sociais e capital social no Twitter. Líbero, São Paulo: 12(24), 81–94. http://seer.casperlibero.edu.br/index.php/libero/article/view/498/472

Sugimoto C. R., Work, S., Larivière, V. & Haustein, S. (2017) Scholarly use of social media and altmetrics: a review of the literature. *Journal of the Association for Information Science and Technology*, 68(9), 2037–2062. DOI:   https://doi.org/10.1002/asi.23833

Van Honk, J. & Costas, R. (2016, September). *Integrating context in Twitter metrics*: preliminary investigation on the possibilities of hashtags as an altmetric resource. Paper presented at the Altmetrics16 Workshop, Bucharest, Romania.

van Raan, A. F. J. (2004). Sleeping Beauties in science. *Scientometrics*. 59(3): 467–472. DOI: https://doi.org/10.1023/B:SCIE.0000018543.82441.f1

Walter, S., Lörcher, I. & Brüggemann, M. (2019). Scientific networks on Twitter: Analyzing scientists' interactions in the climate change debate. *Public Understanding of Science*, 28(6),1–17. DOI: https://doi.org/10.1177/0963662519844131

Wouters, P., Zahedi, Z., Costas, R. (2018). Social media metrics for new research evaluation. In: Glänzel, Wolfang; Moed, Hank; Schmoch, Ulrich; Thelwall, Mike (eds). *Handbook of Quantitative Science and Technology Research*. Springer

Yu, H., Xiao, T., Xu, S. & Wang, Y. (2019). Who posts scientific tweets? An investigation into the productivity, locations, and identities of scientific tweeters. *Journal of Informetrics*, 13(3), 841–855. DOI: https://doi.org/10.1016/j.joi.2019.08.001

---

[1] ANVISA - is the acronym of the Brazilian National Sanitary Surveillance Agency, a regulatory agency, in the form of a special regime autarchy, linked to the Ministry of Health. The agency exercises sanitary control of all products and services (national or imported) subject to health surveillance, such as medicines, food, cosmetics, sanitizers, tobacco products, medical products, blood, blood products and health services.